\documentclass[preprint]{m}
\usepackage{graphicx}
\usepackage[english]{babel}
\usepackage[utf8]{inputenc}
\usepackage{graphicx}
\usepackage{amsmath}
\usepackage{amsfonts}
\usepackage{amssymb}
\usepackage[colorlinks]{hyperref}
\usepackage[thinlines]{easytable}

\date{}

\renewcommand{\refname}{References}

\begin{document}

\begin{frontmatter}
\title{From chaos to clock in reverberating neural net. Case
study}
\author[1]{A. Vidybida\corref{cor1}}
\ead{vidybida@bitp.kiev.ua}
\ead[url]{http://vidybida.kiev.ua/}
\author[1]{O. Shchur}
\ead{olha.schur@gmail.com}
\cortext[cor1]{Corresponding author}
\address[1]{Bogolyubov Institute for Theoretical Physics,
Metrologichna str., 14-B, Kyiv 03680, Ukraine}

\begin{abstract}
What is the reason for complex dynamical patterns registered from real biological
neuronal networks?
Noise and dynamical reconfiguring of a network (functional/dynamic connectome)
were proposed as possible answers.
In this case study, we report a complex dynamical pattern observed in a simple 
deterministic network of 25 neurons with fixed connectome.
After a short initial stimulation, the network is engaged into a complex
dynamics, which lasts for a long time. 
Eventually, with no external intervention, the dynamics comes to a periodic one
with a short period.
The long transient is positively checked for being chaotic.
We conclude that the complex dynamics observed 
is the output of neural computation performed in the process
of neuronal firings and spikes propagation.
\end{abstract}

\end{frontmatter}

\section[Introduction]{Introduction}

Chaotic dynamics in the brain have been observed for a long time. 
Chaos has been reported in electroencephalography (EEG) during sleep, \cite{Babloyantz1985}, or
olfactory perception \cite{Skarda1987}. 
Further investigations reported chaotic dynamical patterns at all levels
of a brain down to  single cells and their membrane conductances, see 
references in \cite{Korn2003}. 
Chaos is recognized now as a normal state of living organism \cite{Pool1989}.
On the other hand, excessive rhythmic activity in the brain is considered as a pathology
in most cases
and should be corrected. See e.g. \cite{Schiff1994} where a possibility for correction
is reported.

Several mechanisms of electrical, chemical and biological nature
are able to  shape dynamics in a biological neural net, see survey in \cite{Breakspear2017}.
In this report, we describe a complex dynamics in a fully connected deterministic network
 of 25 leaky integrate-and-fire (LIF) neurons placed at
 lattice nodes, Fig. \ref{network}. 
Propagation delays are taken proportional to the interneuronal distances. The
network is initially stimulated with a short sequence of 25 input impulses, each triggering one of
the 25 neurons. The sequence of the triggering moments constitutes a stimulus specificity. After
the initial stimulation, the network runs on its own, without external influence
and with no plasticity. 
A stimulus has
been found which triggers a prolonged seemingly chaotic behavior of the network's
state parameters, such as
voltage of a neuron or interspike intervals. This type of dynamics lasts several orders of magnitude
longer than the longest interneuronal communication delay. After that, the dynamics becomes
periodic with a short period.
In order to analyze the transient observed, we apply several tests for complexity and chaos to it.
These tests are as follows: 0-1 test by Gottwald and Melbourne; permutation entropy;
spectral entropy; sensitive dependence on initial conditions. All tests support the idea that
the initial transient  is chaotic.
  This kind of activity looks like an example of
the transient chaos, \cite{Tel2015}.  
Remarkably, none of the used tests 
was able to predict based on initial chunk of the transient whether the dynamics
will fade or settle on a periodic mode, and if the latter, than how long could it
take. These questions will be addressed in further work.

\section{Methods}

\subsection{Neural network}

\begin{figure}
  \begin{center}
    \includegraphics[width=0.5\textwidth,angle=-0]{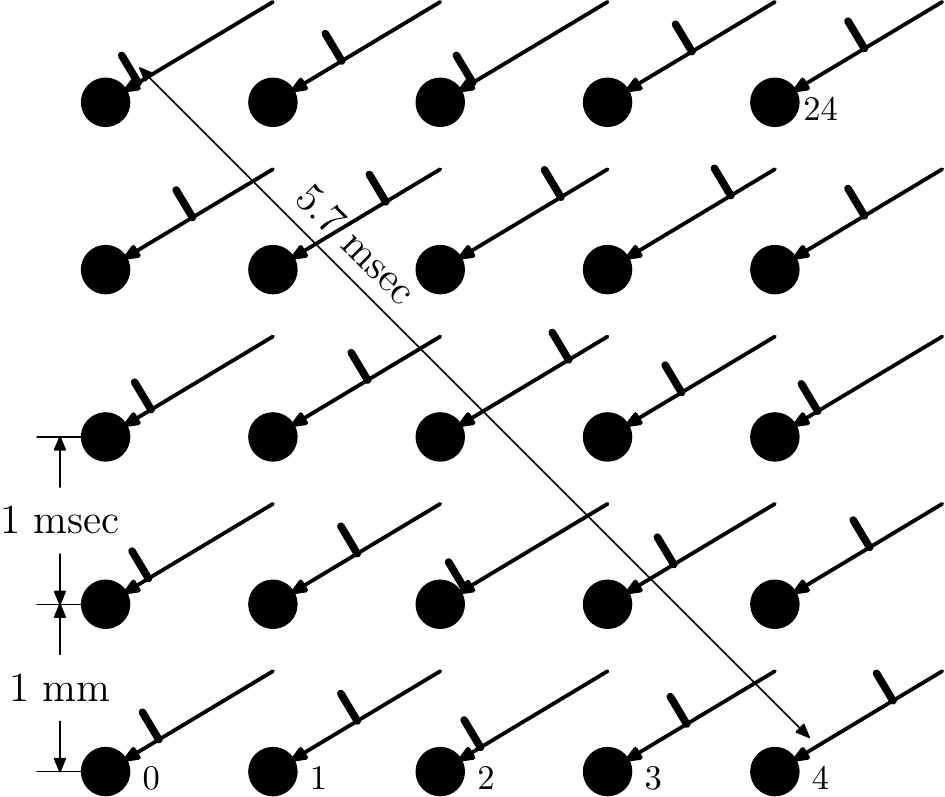}
  \end{center}
\caption{\label{network} Network used for simulations. All neurons are identical leaky integrate-and-fire neurons, with
 threshold voltage $V_{th} = 20$ mV, 
 input impulse height $h = 0.999$ mV,
 membrane time constant $\tau_M = 20$ msec.
 The neurons are simulated with integer numbers as described in
 \cite{Vidybida2019}. Each of 25 bars indicate initial position in time of
 triggering impulse from the stimulus.
}
\end{figure}

The network is similar to that used for numerical simulations in our
previous paper \cite{Vidybida2017a}.
The only difference is the number of neurons (25 instead of 9, see Fig.
\ref{network}). Similar to \cite{Vidybida2017a}, simulation is made with 
time step $dt$ = 0.1 msec.

\subsection{Stimuli}
\begin{figure}
\includegraphics[width=0.45\textwidth,angle=-0]{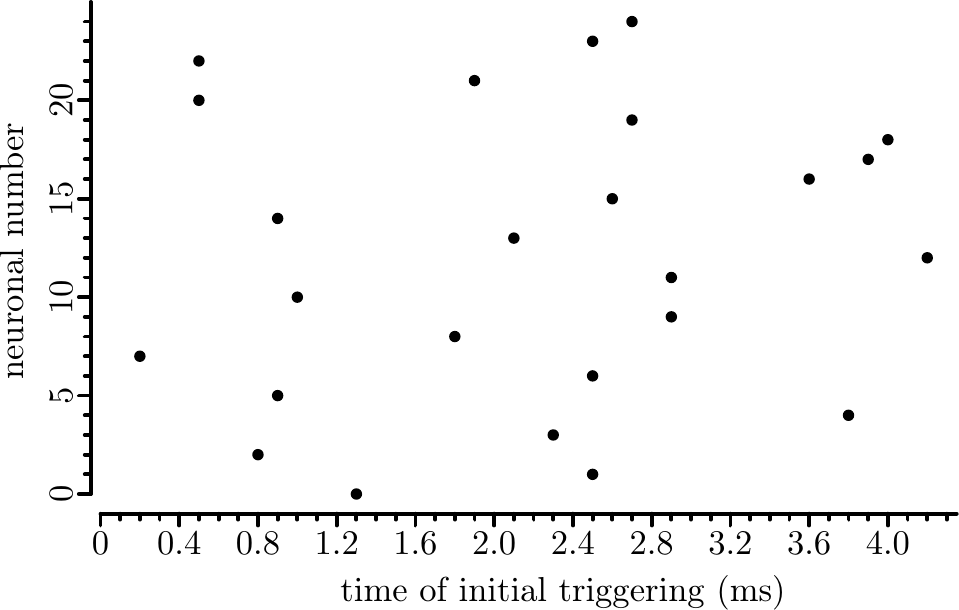}
\hfill
\includegraphics[width=0.45\textwidth,angle=-0]{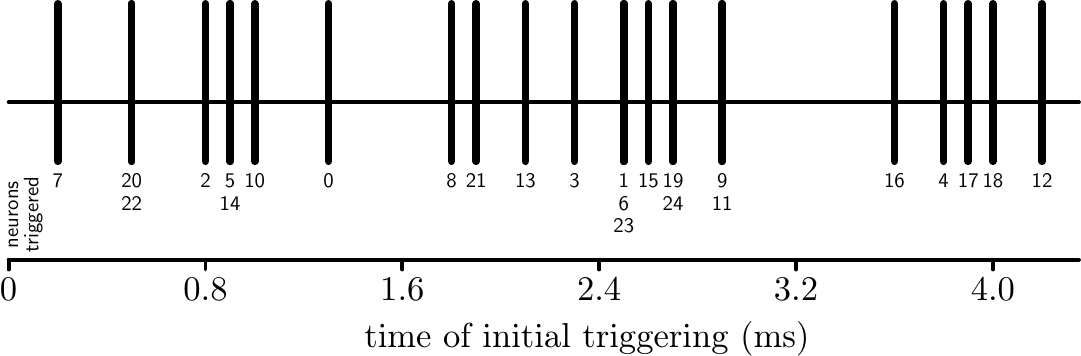}
\caption{\label{stimulus} Initial stimulus used. Left: X-coordinate of a dot indicates the
  moment at which input triggering spike enters corresponding neuron;
  Y-coordinate is the corresponding neuron number.
  Right: the same stimulus displayed as spike train. Numbers under a spike indicate
  neurons triggered by this spike.
  }
\end{figure}

Initial stimulation of the network is performed by applying a sequence of 25
triggering impulses, each for corresponding neuron, see Figs. \ref{network}, \ref{stimulus},
 where the stimulus used in this case study is displayed. 
The stimulus can be depicted as usual as a spike-train, see
Figure \ref{stimulus} (right), but each spike in the train is destined for its own neuron.
So, due to initial stimulation, each neuron obtains single triggering input at
a specified moment. The set of all 25 triggering moments constitutes the
stimulus specificity. After the stimulus is discharged entirely (it takes 4 msec for the
stimulus used), the network's dynamics unfolds on its own.  No further external intervention
is involved and no plasticity or noise is considered.

\subsection{Data acquisition}

During free run after stimulation, the network state at any moment consists of 600 integers characterizing
states of all 600 interneuronal connections, and 25x4 integers characterizing states of
all 25 neurons. The state of a connection is represented by a single integer indicating
 after how many ticks 
the propagating spike will reach its target neuron. If a connection does not convey a spike, 
its state is marked as -1. In a neuronal state, the first two integers represent depolarization
with whole numbers. The other two report whether
the neuron is in {\tt fire} or {\tt empty} state,
see detailes in \cite{Vidybida2019}.
 In order to determine the moment when periodic
regime starts, we add the network state at each tick to a {\tt C++} container 
(actually, the hash of a state) until meet the state, which is already in the container.
The moment of the first appearance of that state is just the moment of entraining onto a
periodic regime. After finding an interesting trajectory, we write it on disc and analyze
by several methods.

\section{Results}

\subsection{Dynamics observed}

\begin{figure}
\includegraphics[width=0.48\textwidth,angle=-0]{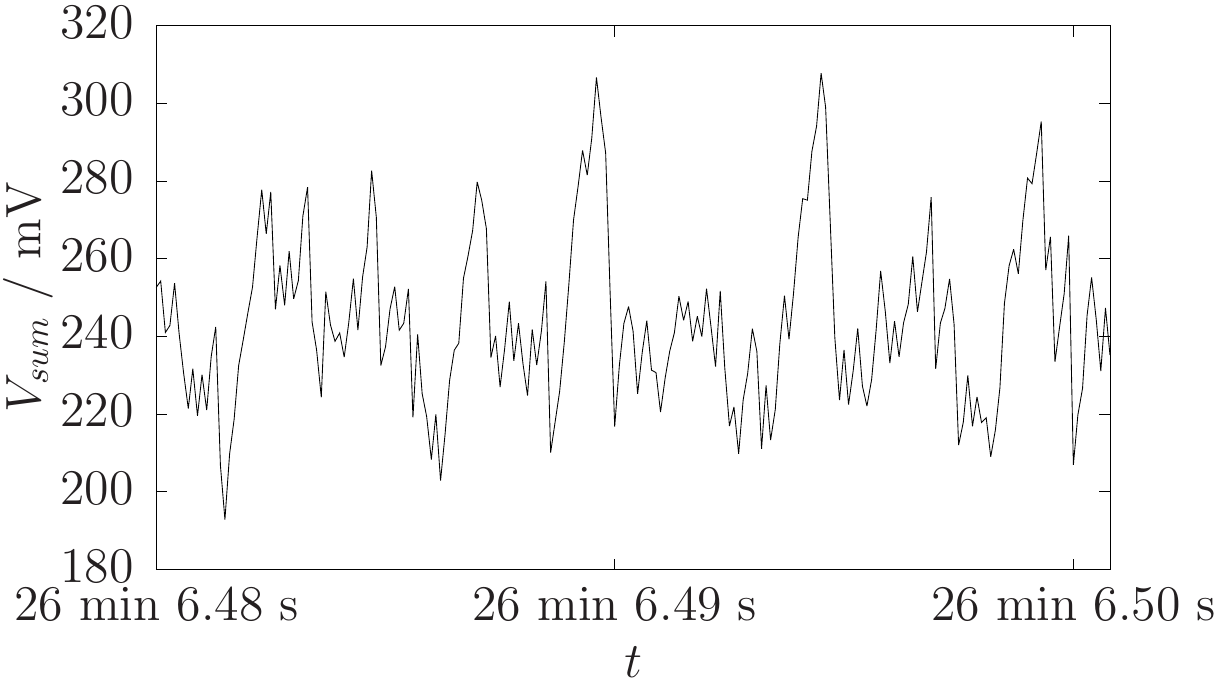}
\hfill
\includegraphics[width=0.48\textwidth,angle=-0]{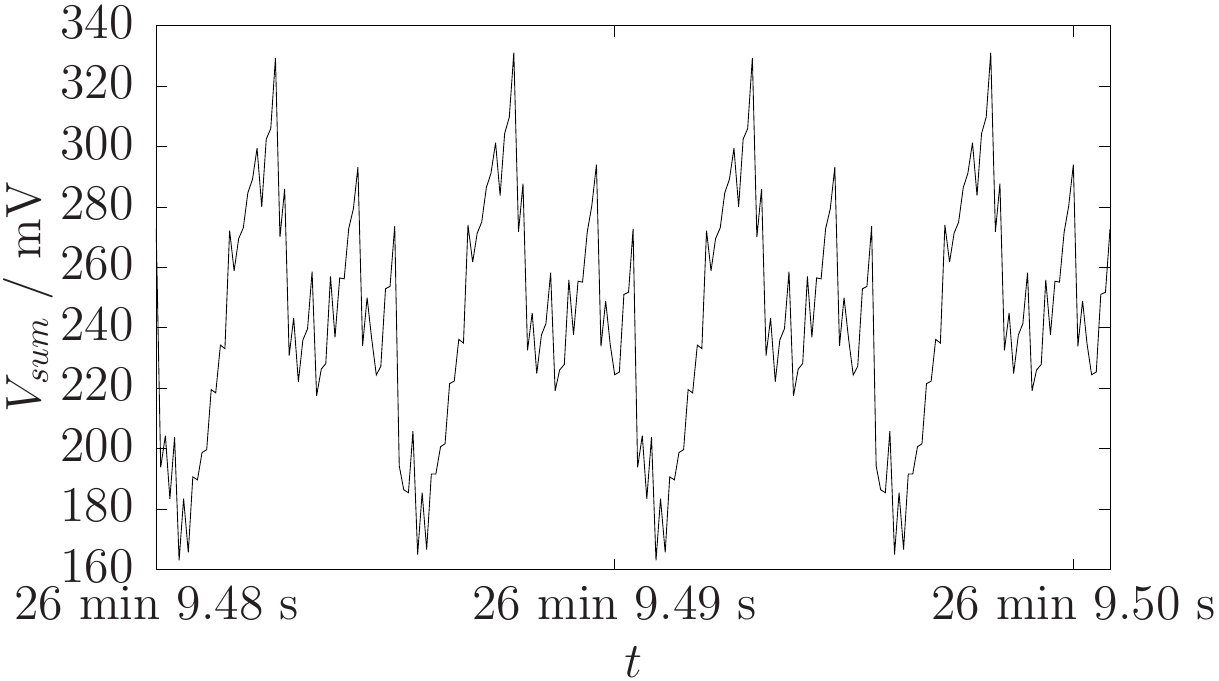}
\caption{\label{chaper}Compound voltage time course after establishing periodic regime (right, two periods long),
and 3 seconds earlier (left).}
\end{figure}

\begin{figure}
\includegraphics[width=0.98\textwidth,angle=-0]{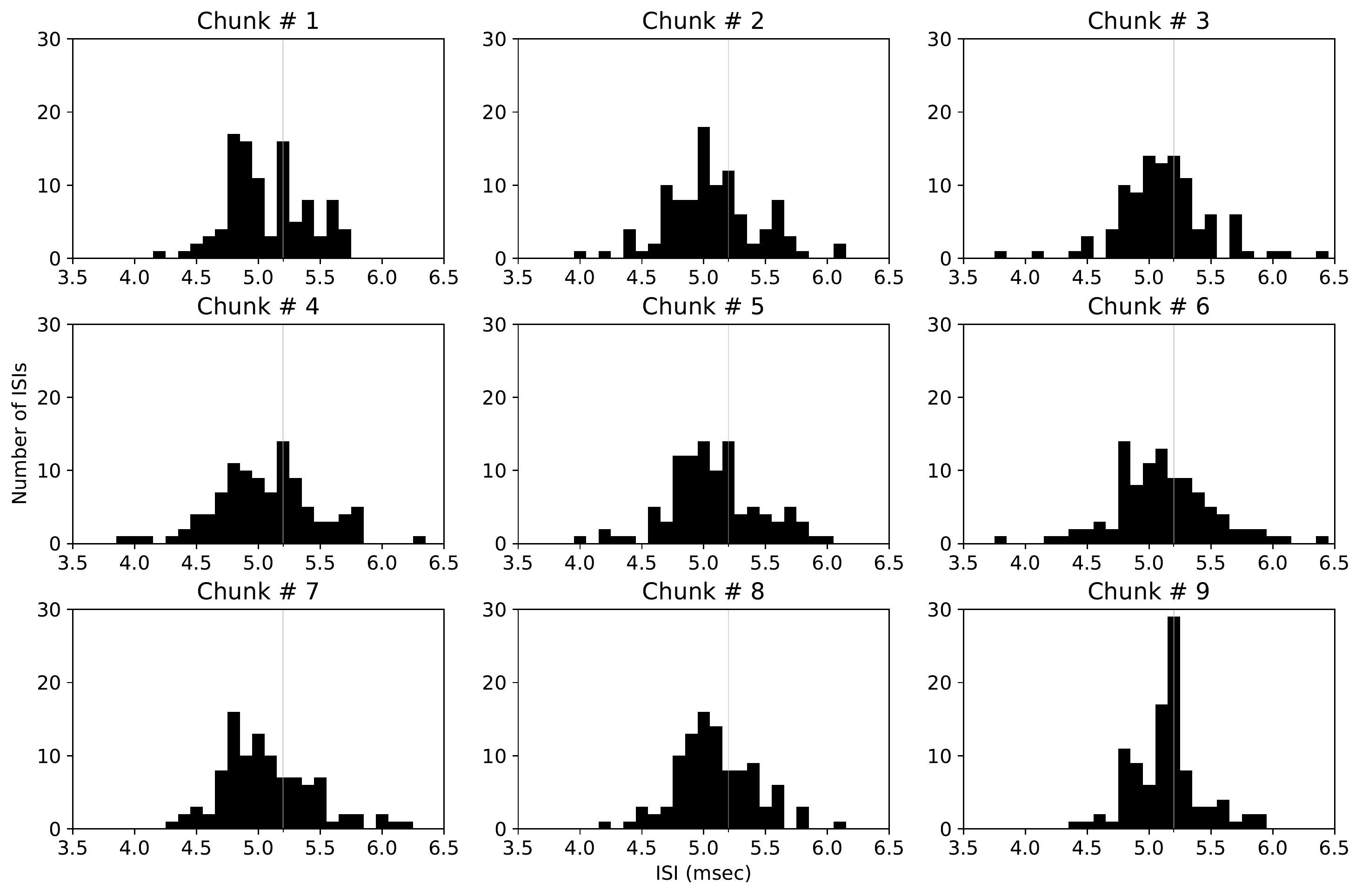}
\caption{\label{ISIhisto}The ISIs histograms for the chunks \# 1-9 for the neuron \# 24.
Here, chunk duration is 0.52 seconds.
Grid lines in all figures correspond to the ISI duration 5.2 msec. 
Note that, in the periodic regime, the neuron \# 24 fires with constant ISI duration 5.2 msec.}
\end{figure}

If stimulated with stimulus displayed in Figs. \ref{network}, \ref{stimulus}, the network demonstrates
the following behavior. During long time (1569.4821 sec = 26 min 9.5 s) the dynamics is seemingly chaotic.
\begin{figure}
\includegraphics[width=0.48\textwidth,angle=-0]{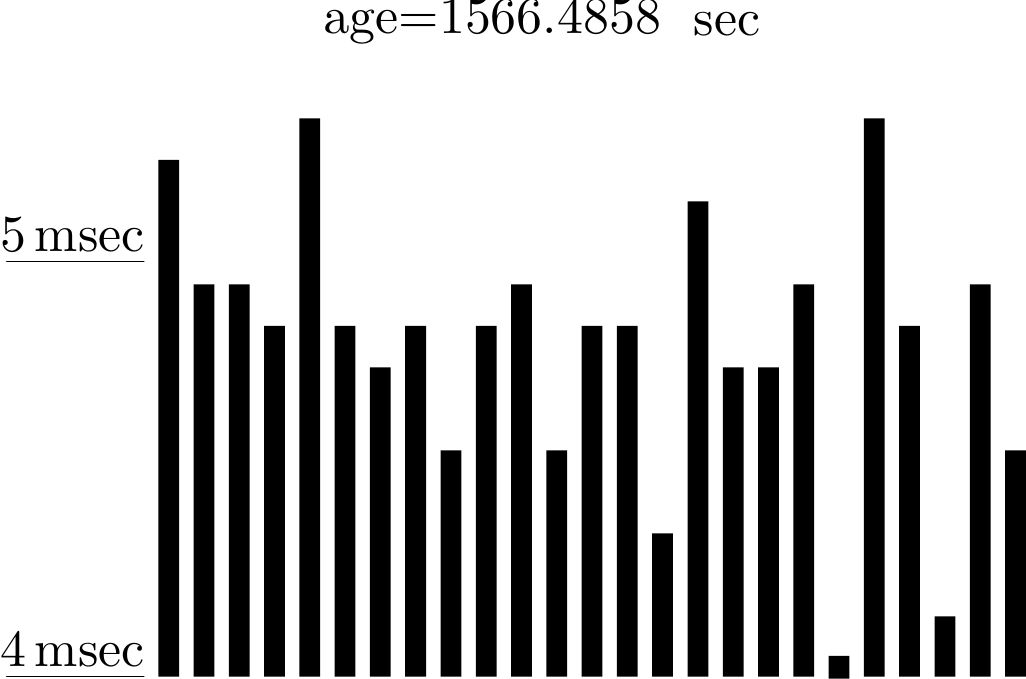}
\hfill
\includegraphics[width=0.48\textwidth,angle=-0]{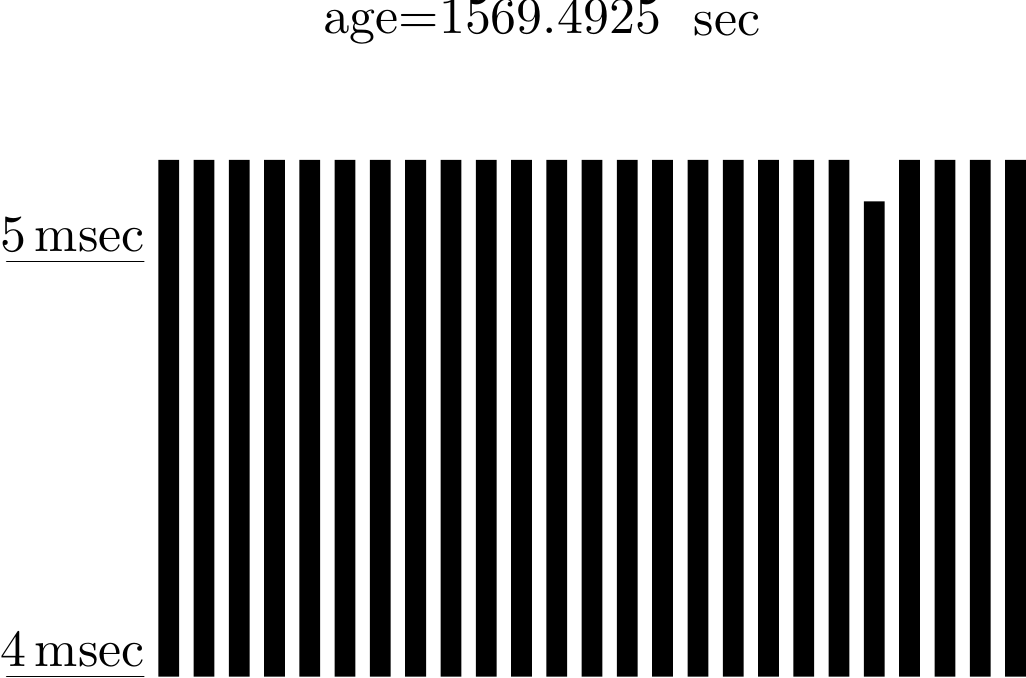}
\caption{\label{ISIsets}Sets of most recent ISIs for all 25 neurons  at time moments specified above as ``age''. Typical ISI set in chaotic (left, 3 sec to periodic regime) and periodic (right) regime.}
\end{figure}
This can be seen from histograms of ISIs for a single neuron 
in different chunks of the transient, Fig. \ref{ISIhisto},
the sets of inter-spike intervals for all 25 neurons, Fig. \ref{ISIsets}(left), and from compound voltage $V_{sum}$ time course, Fig. \ref{chaper}(left), where
\begin{equation}\label{Vsum}
V_{sum}=\sum\limits_{i=0}^{24}v_i\,,
\end{equation}
and $v_i$ is the voltage in neuron \#\,$i$.
Then, within a short period (less than 3 s) dynamics simplifies and eventually 
turns into periodic with period 10.4 msec.
In the periodic regime, 24 of 25 neurons fire with constant ISI duration 5.2 msec.
One neuron fires ISIs 5.1 and 5.3 msec long alternately, e.g. Fig. \ref{ISIsets}(right).
We consider the duration of transient as relatively long. Indeed, time needed for a spike
to cross the diagonal is 5.7 msec, Fig. \ref{network}. Thus, the diagonal crossing may happen
1569.4821 sec / 5.7 msec = 275348 times before the dynamics settles down to periodic regime.
Such a long transient appears as an independent self-reliant dynamics, which can be analyzed
separately.

\subsection{Analysis}\label{An}

\begin{figure}
  \begin{center}
\includegraphics[width=0.8\textwidth,angle=-0]{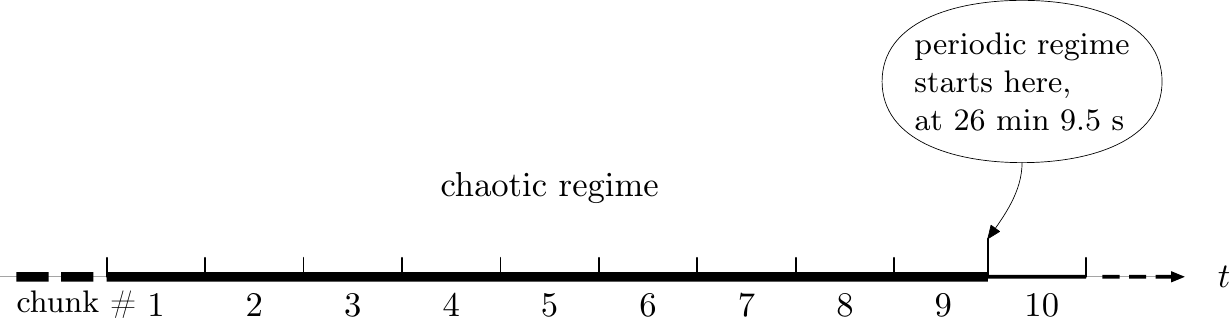}
  \end{center}
\caption{\label{chunks} Chunks used for calculations. Several chunk durations have been tested, see text.
  }
\end{figure}

In order to analyze dynamics in the course of time, we have chosen 10 chunks of the trajectory
in the following way. Each chunk has the same duration. 
Chunks 1..10 follow one another and chunk \# 9 ends just at the beginning of periodic regime.
At this same moment chunk \#10 starts. see Fig. \ref{chunks}.
Three different durations of chunks have been considered, namely, 5200 ticks (50 periods) and
15600 ticks (150 periods), as well as 1743869 ticks (the first nine chunks cover the transient part of
 the trajectory entirely).
As a parameter to analyze we take voltage at a single neuron, $v_i$, $i=0,\dots,24$,
(several neurons have been tested) and the sum of all 25 voltages, $V_{sum}$, (\ref{Vsum}),
which is the analog of the local field potential. The results are similar.
Below, we present results for $V_{sum}$.

\subsubsection{0-1 test for chaos}

Here we use the binary test for chaos
 proposed by G. A. Gottwald and I. Melbourne, \cite{Gottwald2009}.
In this test, a sequence of data
\begin{equation}\label{row}
\mathsf{V} = \{V_0,V_1,\dots, V_{N-1}\}
\end{equation}
can be checked for being chaotic or regular.
In our case $V_i =V_{sum}(i)$, where $i=0,1,2,\dots N-1$, is a tick number
within a chunk and $N$ is the chunk length. 
$\mathsf{V}$
is considered as being either regular or chaotic depending on the behavior of auxiliary
two-dimensional trajectory 
\begin{equation}\label{pqtr}
(p_c(n),q_c(n)),\quad n=0,1,\dots,N-1
\end{equation}
constructed from $\mathsf{V}$ as follows:
\begin{equation}\label{pqdef}
p_c(n)=\sum\limits_{0\le i\le n}V_i\cos(i\cdot c),\quad
q_c(n)=\sum\limits_{0\le i\le n}V_i\sin(i\cdot c)\,.
\end{equation}
Here $c$ is some real number. A set of different $c$ values in the interval (0, $\pi$) 
are considered. If the mean square displacement of $(p_c(n),q_c(n))$ asymptotically 
grows linearly with $n$,
than $\mathsf{V}$ is considered chaotic, otherwise if $(p_c(n),q_c(n))$ stays in a bounded
domain, than $\mathsf{V}$ is regular.
The asymptotic mean square displacement is defined in \cite[Eq. (2.1)]{Gottwald2009} as follows:
\begin{equation}\label{Mcn}
M_c(n)=\lim\limits_{N\to\infty}
\frac{1}{N}\sum\limits_{0\le k<N}
\left((p_c(k+n)-p_c(k))^2+(q_c(k+n)-q_c(k))^2\right),
\end{equation}
which expects infinite number of points in $\mathsf{V}$. In our case, we have a finite number of
points from the end of stimulation to the entrainment onto a periodic regime. Moreover, we 
analyze dynamics in 9 consecutive chunks before the periodic regime starts, see Fig. \ref{chunks}.
This limits possible value of $N$ in (\ref{Mcn}) by 1743869. Thus, the modified for our case mean
square displacement in each chunk is calculated as follows:
\begin{equation}\label{Mcn1}
M_c(n)=
\frac{1}{N_1}\sum\limits_{0\le k<N_1}
\left((p_c(k+n)-p_c(k))^2+(q_c(k+n)-q_c(k))^2\right),
\end{equation}
where $n=0,1,\dots,n_1=1000$, and the following condition is satisfied:
$$
N_1+n_1\le N,
$$
where $N$ is the chunk length.

\begin{figure}
\includegraphics[width=0.5\textwidth,angle=-0]{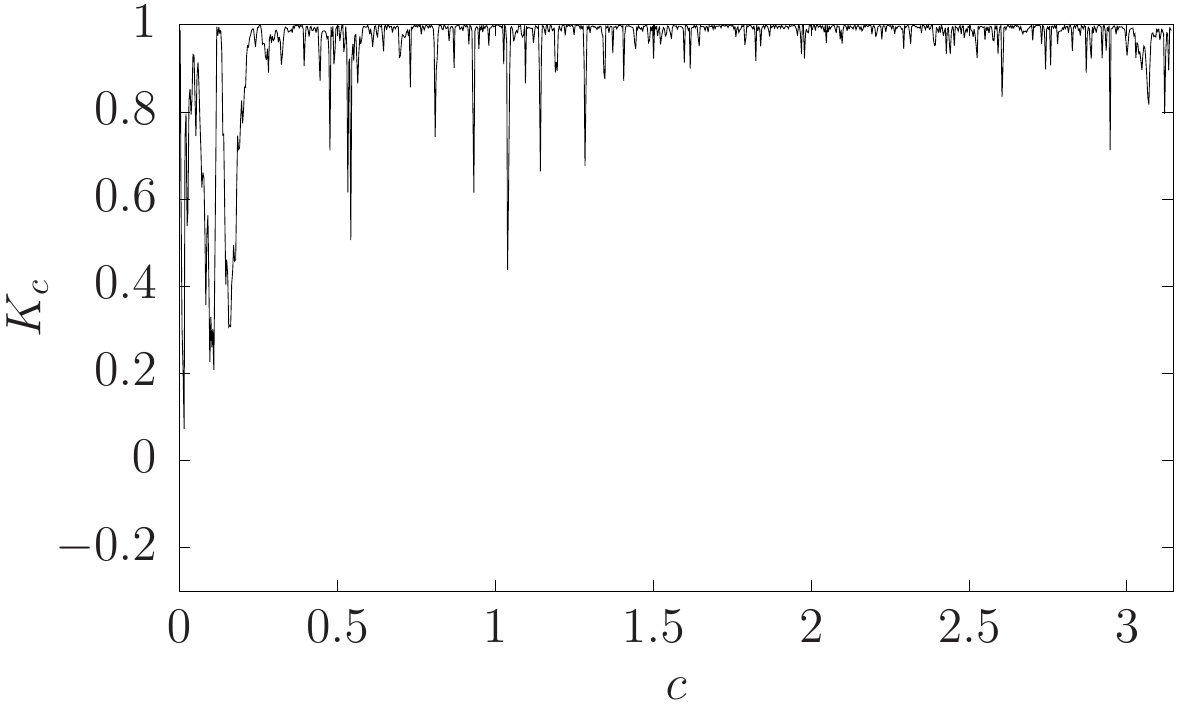}
\hfill
\includegraphics[width=0.5\textwidth,angle=-0]{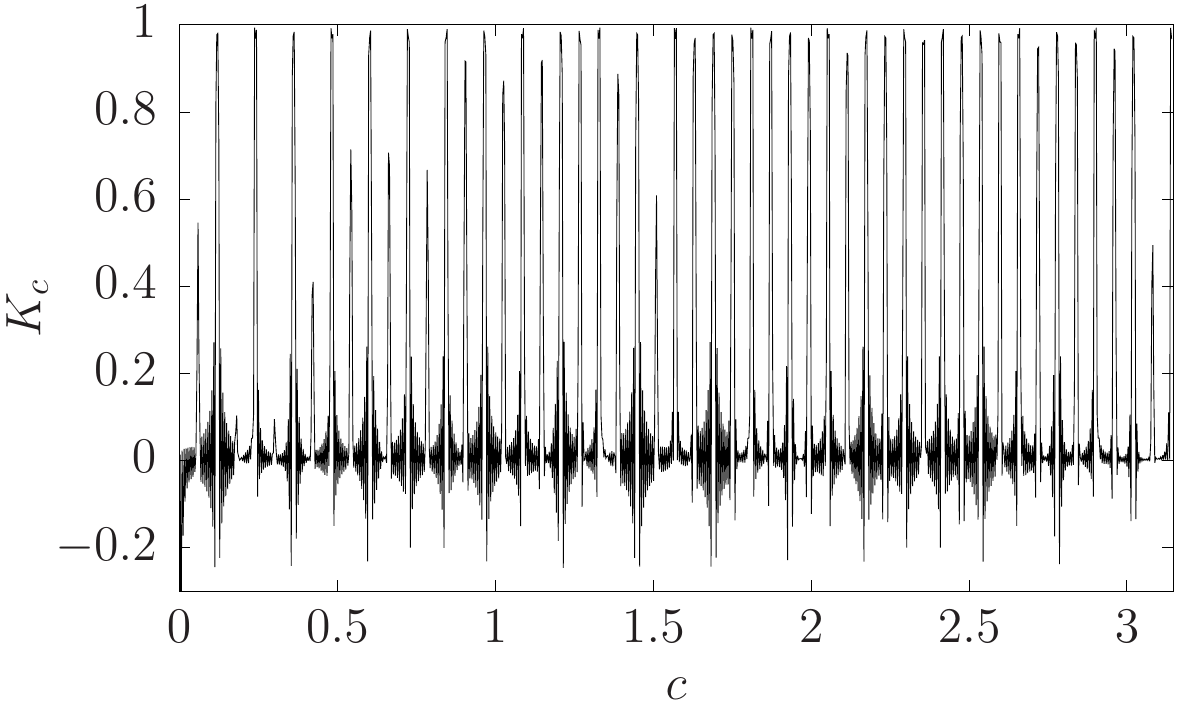}
\caption{\label{Kc4-9}Behavior of correlation coefficient $K_c$ as a function of $c$-value. Left: chunk \#9, right: chunk \#10 (periodic regime).  Here, chunk duration is 0.52 seconds; 1000 equidistant values
of $c\in [0;\pi]$ were used for calculations.}
\end{figure}

Mean square displacement $M_c(n)$ oscillates with $n$, which  impairs convergence.
The oscillations can be subtracted from 
$M_c(n)$ as it is proposed in \cite[Eq. (2.3)]{Gottwald2009}. The resulting quantity $D_c(n)$
is checked for linear growth by correlation method, see \cite[Sec. 3.2]{Gottwald2009}.
The correlation coefficient $K_c$ between sequences $\{1,2,\dots,n_1\}$ and
$\{D_c(1),D_c(2),\dots,D_c(n_1)\}$ is calculated for different values of $c$,
and median in the obtained $K_c$ set was found. Resulting median for the first nine chunks of the
trajectory is close to 1 (see Fig. \ref{Kc4-9}, Table \ref{0-1med}) qualifying those chunks
as chaotic.
 All three values for $N$ mentioned above in Sec. \ref{An} were tested with $n_1=1000$. 
The results are similar. 

\begin{table}
{\small 
\begin{TAB}(@,0.7cm,0.7cm)[3pt]{|c|c|c|c|c|c|c|c|c|c|c|}{|c|c|}
chunk number & 1 & 2 & 3 & 4 & 5 & 6 & 7 & 8 & 9 & 10 \\
median of $K_c$ & 0.992 & 0.994 & 0.993 & 0.994 & 0.993 & 0.994 & 0.993 & 0.994 & 0.992 & 0.026 \\
\end{TAB}
}
\caption{\label{0-1med}Result of 0-1 test for chaos in each chunk. Here, chunk duration is 0.52 seconds.}
\end{table}

\subsubsection{Permutation entropy}
\begin{figure}
\center
\includegraphics[width=0.75\textwidth,angle=-0]{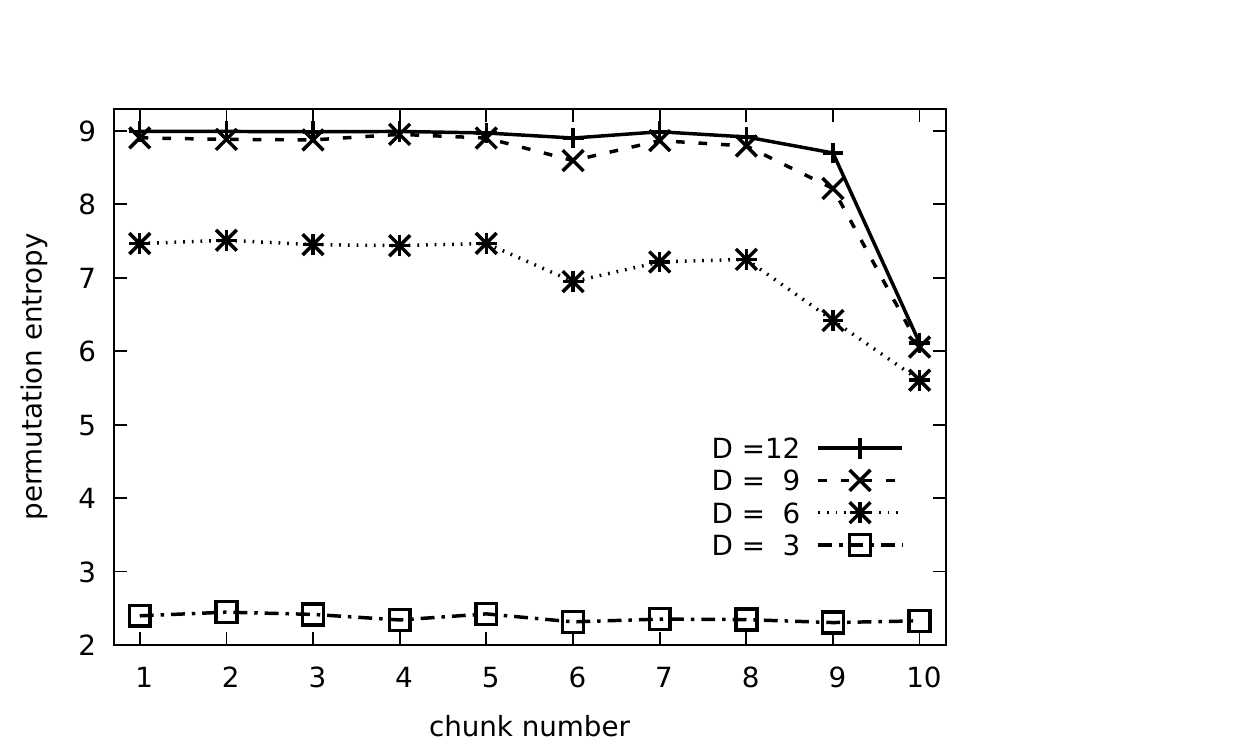}
\caption{\label{PE}Permutation entropy for different embedding dimensions $D$
calculated for $V_{sum}$ in chunks shown in Fig. \ref{chunks}. 
Here, chunk duration is 0.52 seconds. The embedding delay $\tau=1$.}
\end{figure}

Complexity of trajectory $V_{sum}(t)$ in different chunks was also analyzed by calculating 
permutation entropy in each chunk.
The permutation entropy method is proposed for estimating complexity of trajectories of a dynamical 
system, see \cite{Bandt2002}. In order to apply this method to a sequence of data (\ref{row}),
one needs to chose an embedding dimension $D>1$ and create a sequence of embedding vectors
$\mathbb{V} = \{\mathbf{V}_0,\mathbf{V}_1,\dots, \mathbf{V}_{N-D}\}$, $\mathbf{V}_i\in \mathbb{R}^D$, $i=0,1,2,\dots N-D$,
where $\mathbf{V}_i= (V_i,V_{i+1},V_{i+2},\dots, V_{i+D-1})$. An additional parameter of the embedding 
procedure is delay, which we choose 1 here. Further step in the method is to find for each
$\mathbf{V}_i$ a permutation $\pi_i$, which arranges its components in ascending order. 
$\pi_i$ is called the order pattern of $\mathbf{V}_i$.
Having a sequence of order patterns $\Pi=\,\{\pi_0,\pi_1,\dots,\pi_{N-D}\}$, we calculate the probability 
$p_i$ of any of $D!$ possible patterns by dividing the number of its occurrences in $\Pi$
by the total number of elements in $\Pi$. 
The permutation entropy of $\mathbb{V}$ is the Shannon entropy of the probability
distribution $p(\pi_i)$:
\begin{equation}\nonumber 
S(\mathbf{\mathbb{V}})\equiv S(\Pi)=-\sum\limits_{i=0}^{M-1}p(\pi_i)\log(p(\pi_i)),
\end{equation}
where $M$ is the number of different permutations in the $\Pi$.
In this work we use a modification of this method, arithmetic entropy, which is exempt of
combinatorics, see \cite{Vidybida2020}. 
 Both methods deliver the same value for entropy provided that in any 
embedding vector all $D$ components are different, see \cite[Theorem A.1]{Vidybida2020}. In our case,
 we registered voltages with nine decimal places, therefore equalities are improbable.
 Additionally, 
in the program we developed, a check for equality between components of embedding vector
  was introduced, and we observed no equal components in $\mathbf{V}_i$.
 The resulting entropy values are shown in Fig. \ref{PE}.
From these data we see that the trajectory $V_{sum}(t)$ has high complexity, 
roughly the same for the first 8 chunks. In the ninth chunk, complexity decreases slightly,
and falls to lower values at the tenth chunk. In the periodic regime, the permutation entropy
is still considerably high. This can be understood having in mind that the periodic regime
itself is still quite complex, see Fig. \ref{chaper} (right). 
For embedding dimensions considered, except of $D=3$, periodic part of the trajectory produces
enough different order patterns.

\subsubsection{Spectral entropy}

Spectral entropy is one of nonlinear dynamics and chaos theory methods used
to analyze EEG signals, \cite{Rodriguez-Bermudez2015}.
Different entropy measures can be used to detect
epileptic seizure on EEG. In a review \cite{Acharya2015},
it was concluded that the spectral entropy is among the best entropy measures to
perform in this task.  Also different entropy measures,
including the spectral entropy,
are used to classify emotions from EEG in a brain-computer interface, see 
recent review \cite{Patel2021}. Additionally, applied to local field potential measurements,
correlation of time varying spectral entropies  is used  to
detect synchrony in neural networks \cite{Kapucu2016}.
Therefore, we have tested the transient with this metod.

Firstly, different chunks of the trajectory $V_{sum}(t)$ were analyzed by 
calculating the spectral power density
(PSD) and subsequently the power spectral entropy (PSE).
The algorithm used to calculate PSD is as follows.
In the current work, during the course of computer simulation, the sum of voltages of all 25 neurons $V_{sum}$
(\ref{Vsum}) was sampled at discrete times with step $dt$. For one chunk,  sampled $V_{sum}$ is given by $\mathsf{V}$ as in (\ref{row}) with $N=5200$, where $N$ is the size of a chunk. Firstly, one needs to calculate the discrete Fourier
transform of the sequence $\mathsf{V}$:
\begin{equation}\label{row_fourier}
\mathsf{\tilde{V}} = \{\tilde{V}_0,\tilde{V}_1,\dots, \tilde{V}_{N-1}\}.
\end{equation}

 Then, from the discrete Fourier
transform of $\mathsf{V}$, the 
PSD  at the frequency $f_k = \frac{k}{N dt} $, $k=1,\ldots,N/2$, can be calculated 
using the following expression:
\begin{equation}
PSD(f_k) = \dfrac{|\tilde{V}_k|^2 dt}{N} .
\end{equation}

\begin{figure}
  \begin{center}
\includegraphics[width=\textwidth]{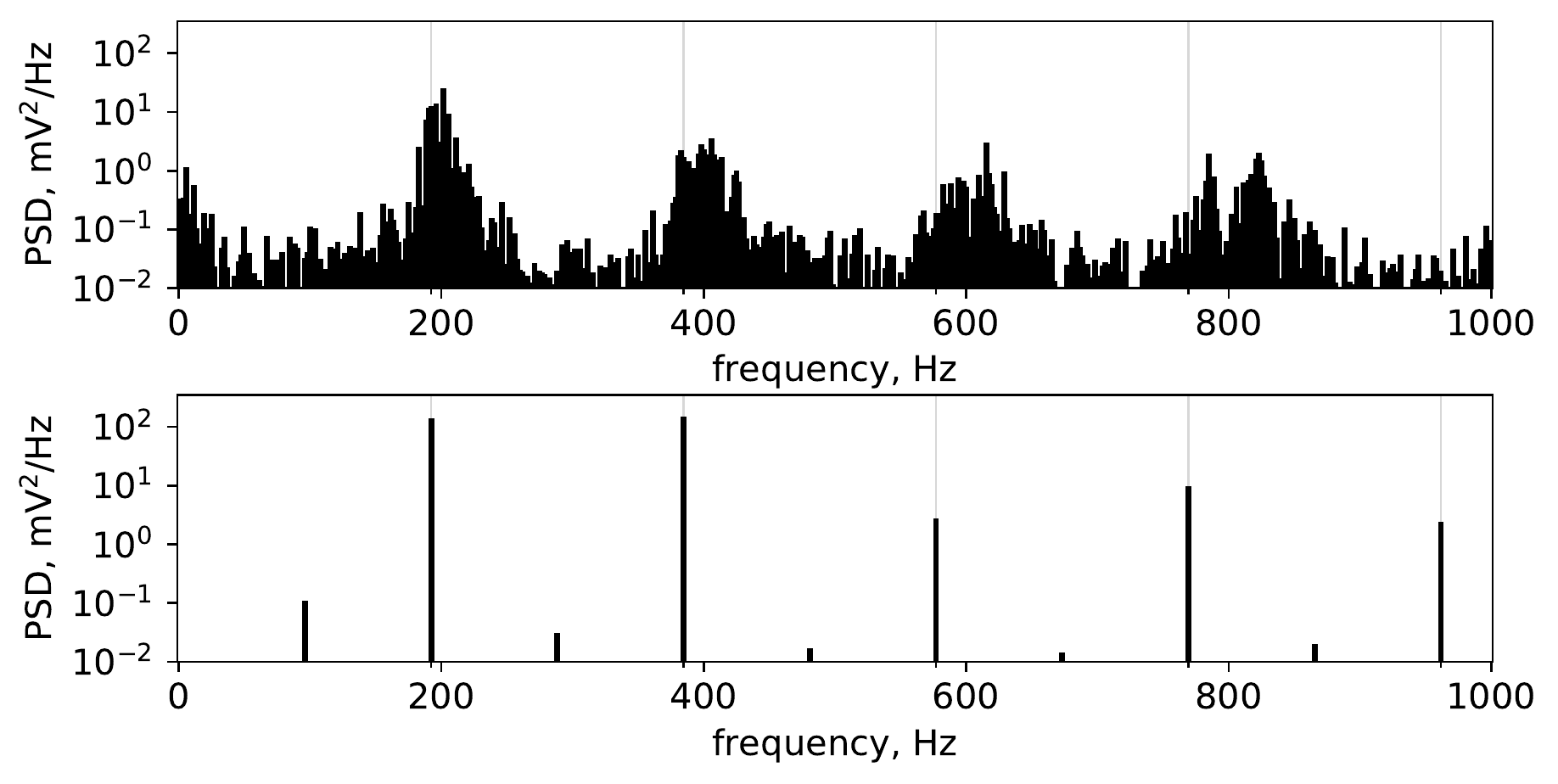}
  \end{center}
\caption{\label{stimulus-why?} The power spectral density (PSD) of the sum of voltages of all 25 neurons $V_{sum}$ (\ref{Vsum}) calculated on the chunks \# 1 (upper panel)  and \#10 (lower panel). Only frequencies up to 1 kHz are shown. Grid lines in both figures correspond to the frequency 1/5.2 kHz and its harmonics.
  }
\end{figure}

The PSD was calculated for all ten chunks. For the chunks \#1 and \#10, the PSD is depicted on Fig.
\ref{stimulus-why?}. Note that on the chunk \#10 (periodic activity, lower panel of Fig.
\ref{stimulus-why?}) the frequency 1/5.2 kHz and its harmonics have the most power, while on the chunk \#1 (upper panel of Fig.
\ref{stimulus-why?}) wide peaks in the vicinity of the frequency 1/5.2 kHz and its harmonics are present.

The PSE, or  Shannon spectral entropy, is an application of 
Shannon entropy expression to the power spectral density components \cite{Acharya2015}.
To calculate the PSE, the PSD is usually normalized with the total power:
\begin{equation}
PSD_{norm}(f_k) = \dfrac{PSD(f_k)}{\sum_{l=1}^{N/2}PSD(f_l)  }.
\end{equation}

Then the power spectral entropy  is given by the following formula:
\begin{equation}
PSE = - \dfrac{1}{\log _2 \left( N/2 \right)} \sum_{k=1}^{N/2} PSD_{norm}(f_k) \log _2 \left( PSD_{norm}(f_k) \right).
\end{equation}

Note that here the PSE is normalized with $\log _2 \left( N/2 \right)$, which
is the maximal PSE of a white noise having equal intensity at all frequencies.

\begin{figure}
  \begin{center}
\includegraphics[width=0.7\textwidth,angle=-0]{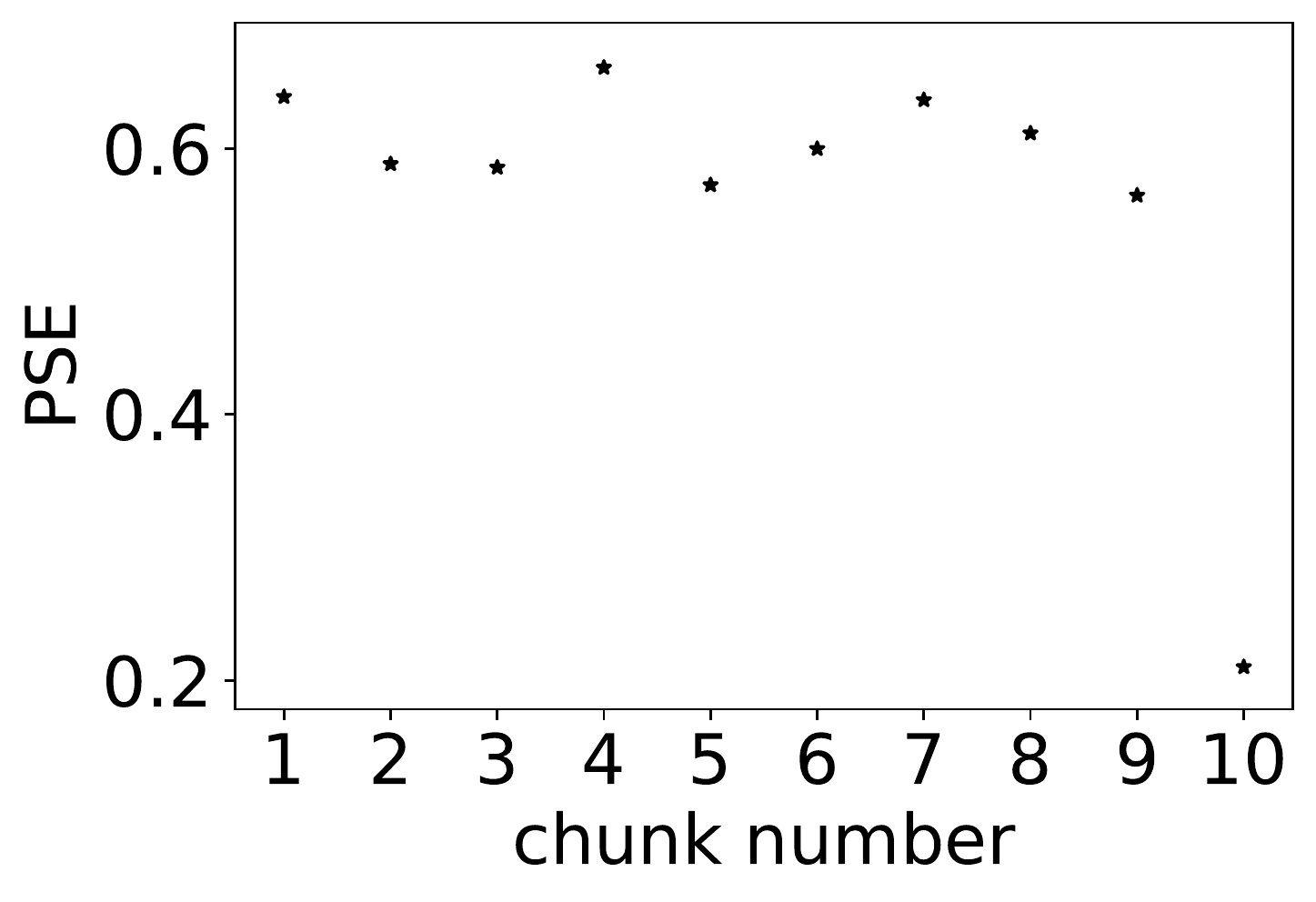}
  \end{center}
\caption{\label{stimulus-why-again?} The normalized power spectral entropy (PSE). X-coordinate of a star indicates the chunk number. 
  }
\end{figure}

The normalized PSE was calculated for chunks \#1-10. The results of calculations are depicted on Fig. \ref{stimulus-why-again?}. The PSE is roughly the same for the chunks \#1-9 during the relaxation to periodic activity, and sharply drops on the chunk \#10 for periodic activity.

\subsubsection{Sensitivity to small perturbations}\label{stabi}
It is known that for chaotic dynamical systems small
perturbations of initial state are able to produce large 
divergence of resulting trajectories.
In our case, initial state is achieved at the end of a stimulus applied
to the standard state with empty neurons and axons.
Naturally, small difference between initial stimuli results in small difference
in the network's initial state. Therefore, and having in mind that a network
is aimed at processing initial input into an output, we consider here small
perturbations of the initial stimulus.
A stimulus specificity is determined by moments of initial
triggering of each neuron, see Fig. \ref{stimulus}. Since we have a finite time step $dt=100\mu s$,
the smallest possible perturbation of the stimulus can be prepared by shifting triggering moment
of a single neuron by $\pm dt$. This gives 50 different stimuli characterized by the smallest possible
deviation from the initial one. Each perturbed stimulus was applied and resulting dynamics
was analyzed. The summary is presented in the Table  \ref{stab}. We see that some of the perturbed stimuli
cause dynamics which ends up with periodic regime with other period than the unperturbed one.
This can be treated as sensitive dependence on initial conditions/stimulus,
which characterizes chaotic dynamics.

  \begin{table}
  \begin{TAB}(@,0.6cm,0.7cm)[5pt]{|c|c|c|c|}{|c|c|c|c|c|c|c|c|}  
  Period, ms & \vbox{\hbox{\small Number of spikes}\hbox{\small each neuron emits}
  \hbox{\raisebox{-1.6mm}[1mm][0.5mm]{\small per period}}} & Number of cases & Relaxation time, s \\
  \vtop{\hbox{\phantom{activit}0}\hbox{(activity fades)}} &0& 2 & 0.08,  0.09\\
  9.6 &2& 3 & 60 - 436\\
  9.7 &2& 2 & 30, 257\\
  9.8 &2& 2 & 344\\
  10.0 &2& 2 & 368\\
  10.4 &2& 35 & 26 - 1569  (26 min)\\
  41.6 &8& 5  & 36 - 722\\
  \end{TAB}
  \caption{\label{stab}
   Result of checking stability with respect to perturbation of the initial
   stimulus.}
  \end{table}

\section{Conclusions and discussion}

In this report, in a simple deterministic neural network simulated in a PC, 
we have observed a remarkable dynamics 
in which, after a short initial stimulation, a very long chaotic transient
comes to a periodic activity. This is not the first observation of this type
in a simulated neuronal network, see, e.g. \cite{Zumdieck2004,Zillmer2009}.
In our case, external intervention is given as a short initial stimulus
after which dynamics runs by itself. In the previous observations external
input is constantly present. 

We have checked the transient observed for being chaotic by several known methods.
In this connection it should be noted that standard definition of chaos
is made for systems with continuous trajectories in a metric space
(see a discussion of exact Devaney's definition of chaos in \cite{Banks1992}).
In our case, trajectories 
are not continuous: each neuronal firing produces a discontinuity.
Also, due to the numerical modeling, our trajectories are confined in a finite set of
state points. Therefore, straightforward application of chaos definition 
made for continuous systems does not make sense in our case. 
The methods applied here (0-1 test for chaos, spectral and permutation entropy
estimates of complexity) do not expect continuity and can be applied to 
discontinuous trajectories. 
The appealing property of chaos known as sensitive dependence on initial conditions
expects a possibility to consider infinitesimally small perturbations
of initial conditions, see \cite{Banks1992}. In our case, the smallest possible
perturbations are finite. Some of these perturbations result in trajectories
differing qualitatively from the unperturbed one, see Table \ref{stab}.

The reason for a neural network to reproduces chaos can be various,
 see \cite{Faure2001} for discussion. 
Our modeling algorithm operates in whole numbers excluding possibility
of rounding errors.
Also, no noise is considered.
It seems that the only reason for complex behavior in our network is a kind
of neural computation performed due to neuronal firings and resulting
in rearrangement of interspike intervals in accordance with the rules imposed
by the interneuronal communication delays and LIF neuron parameters. 
This is a kind of self-organization in the time domain envisioned by D. M. MacKay,
\cite{MacKay1962}.
The computation ends when the periodic mode is reached or the activity fades.
Different stimuli can cause different periodic modes in the same network.
In our previous work \cite{Vidybida2011,Vidybida2017a} made for smaller networks
similar behavior has been observed, but with short transients.

In this connection a question arises: To  what extent the structure of a neuronal
network (``connectome'', \cite{Sporns2005}) determines its function?
 From the point of view of physics the answer should be the following: completely.
 But it appeared quite difficult to map brain structure to function 
(if function is considered as a concrete dynamics evoked by a concrete stimulus) based
 exclusively on the connectome. The idea of a ``dynome'' was proposed as some
 additional rules governing the dynamics, \cite{Bargmann2013,Kopell2014}.
In parallel, concepts of  ``functional connectome'',
 \cite{Biswal2010}, or  ``dynamic connectome'',  \cite{Kringelbach2020}, 
were proposed. In these concepts, connections in the brain can be 
functionally/dynamically reconfigured
depending on the cognitive task, or neuromodulator presence. 
As a result, complex brain dynamics are generated.
The case reported here, see also \cite{Vidybida2011,Vidybida2017a}, 
demonstrates that complex, stimulus dependent dynamic
repertoires can also be generated 
without external intervention
in a deterministic recurrent network with fixed connectome
exclusively due to the numbers game performed in the process of neural computation.
\bigskip

\noindent
{\small\bf Acknowledgments.} {\small 
This work was supported by the Programs of Basic Research
of the Department of Physics and Astronomy of the National Academy of Sciences
of Ukraine “Noise-induced dynamics and correlations in nonequilibrium systems”, № 0120U101347.
AV thanks to V. Mochulska for a good literature survey during her undergraduate study
in Taras Shevchenko National University of Kyiv.}
\bigskip
\vspace*{1cm}




\end{document}